# Testing holographic duality in hyperbolic lattices

Jingming Chen[1], Feiyu Chen[2], Linyun Yang[1], Yuting Yang[3], Liren Chen[1], Zihan Chen[1], Ying Wu[4], Yan Meng[1], Bei Yan[1], Xiang Xi[1], Zhenxiao Zhu[1], Minqi Cheng[1], Gui-Geng Liu[5], Perry Ping Shum[1], Hongsheng Chen[6], Rong-Gen Cai[7], Run-Qiu Yang[2,*], Yihao Yang[6,†], Zhen Gao[1,‡]

[1]State Key Laboratory of Optical Fiber and Cable Manufacturing Technology, Department of Electrical and Electronic Engineering, Guangdong Key Laboratory of Integrated Optoelectronics Intellisense, Southern University of Science and Technology, Shenzhen 518055, China.

[2]Center for Joint Quantum Studies and Department of Physics, School of Science, Tianjin University, Tianjin 300350, China.

[3]School of Materials Science and Physics, China University of Mining and Technology, Xuzhou 221116, China.

[4]School of Physics, Nanjing University of Science and Technology, Nanjing 210094, China.

[5]Department of Electronic and Information Engineering, School of Engineering, Westlake University, Hangzhou 310030, China.

[6]Interdisciplinary Center for Quantum Information, State Key Laboratory of Extreme Photonics and Instrumentation, ZJU-Hangzhou Global Science and Technology Innovation Center, College of Information Science and Electronic Engineering, ZJU-UIUC Institute, Zhejiang University, Hangzhou 310027, China.

[7]Institute of Fundamental Physics and Quantum Technology, Ningbo University, Ningbo, 315211, China.

[*†‡]Corresponding author Email: gaoz@sustech.edu.cn (Z.G.); yangyihao@zju.edu.cn (Y.H.Y.); aqiu@tju.edu.cn (R.-Q.Y)

**Abstract:**

The celebrated holographic duality posits a correspondence between a quantum gravity in a bulk spacetime and a quantum field theory (QFT) defined on its lower-dimensional boundary. This duality not only offers deep insights into the enigmatic nature of quantum gravity but also provides an efficient methodology for studying strongly correlated systems. However, despite its profound significance in modern physics, holographic duality remains a conjecture, and further experimental exploration is highly sought after. Here, we present the first experimental test of holographic duality between a three-dimensional bulk gravity and a two-dimensional boundary QFT using hyperbolic lattices. By experimentally measuring the classical scalar field propagator in hyperbolic circuits, we reproduce the equal-time two-point correlation function of the dual boundary conformal field theory (CFT), verifying its exponential dependence on the boundary separation and the conformal dimension-scalar mass relation. Furthermore, by leveraging the two-point correlation function, we reconstruct the entanglement entropy for a boundary CFT subsystem, confirming that it follows the Ryu-Takayanagi formula. These results constitute the first direct experimental evidence that quantum properties of the QFT can be holographically reproduced through its dual classical field in curved space. This heuristic experimental effort opens a new avenue for in-depth investigations on the holographic duality and extensive exploration of quantum-gravity-inspired phenomena in classical systems.

**Introduction:**

The gravitational holographic duality (*1, 2*) stands as one of the most profound insights in modern physics, asserting that the universe may fundamentally resemble a hologram: all the information contained in a volume of space can be fully encoded on its lower-dimensional boundary. More concretely, this duality postulates that a $d$+1-dimensional weakly coupled quantum gravity is equivalent to a strongly coupled, nongravitational quantum field theory (QFT) on its $d$-dimensional boundary. This revolutionary framework has not only significantly deepened our understanding of the quantum origin of spacetime (*3-5*) and the nature of black hole entropy/information (*5-7*) but also provided a valuable analytical framework for studying strongly interacting systems in both nuclear (*8*) and condensed matter physics (*9*).

Despite its monumental significance, holographic duality remains a conjecture. Its direct experimental verification has long been considered a formidable challenge, if not impossible, as it necessitates the independent measurement of observables in a strongly correlated quantum field and in a real-world, higher-dimensional gravity system to examine their consistency. Remarkably, a key insight provides a way forward: this duality also implies a classical-quantum correspondence, namely the leading-order effects of the strongly coupled boundary QFT can be holographically described by a higher-dimensional, weakly coupled bulk field theory in a curved space whose dominant contributions are governed by classical dynamics in a fixed gravity background, as exemplified by the holographic correlation function (*10, 11*) and holographic superconductor (*12, 13*). This motivates a paradigm-shifting experimental approach to testing the holographic duality, namely by examining whether a classical field in a suitably designed curved space could holographically reproduce certain quantum features of the dual boundary QFT.

On the other hand, recent experimental breakthroughs in simulating scalar field dynamics (*14, 15*) and topological states (*16-20*) in hyperbolic lattices (*14-32*), discretized regularizations of two-dimensional (2D) curved spaces with constant negative curvature, have exemplified an effective approach for the discrete implementation of classical fields in 2D hyperbolic space. This versatile hyperbolic lattice system serves as an ideal platform for investigating holographic duality following the aforementioned strategy. Nevertheless, two fundamental limitations persist in such setups. First, a 2D hyperbolic space is holographically dual to a one-dimensional (1D) QFT, a quantum mechanics system without spatial extent, thereby precluding the study of physical quantities that depend on the spatial characteristics of 2D or higher-dimensional QFTs, such as holographic entanglement entropy. Second, existing hyperbolic lattices are incapable of emulating 2D hyperbolic spaces with non-trivial geometric topologies, such as wormholes or black holes.

Here, we overcome these challenges and report the first experimental test of holographic duality between a three-dimensional (3D) bulk gravity and a 2D boundary QFT using hyperbolic circuits, i.e., electrical circuit realizations of hyperbolic lattices. We transcend the limitations of hyperbolic lattices by introducing laboratory time as the third dimension onto both the conventional type-I and the newly discovered type-II hyperbolic lattices to extend their dimensionality, thereby modeling 3D gravity backgrounds knowns as pure Euclidean Lifshitz space and Euclidean Lifshitz wormhole in the large dynamical exponent limit, which serve as the gravity duals of 2D nonrelativistic conformal field theories (CFTs). Through time-resolved pulse measurements, we experimentally identify distinct spatiotemporal geodesics in hyperbolic circuits, both without and with an analogue 3D wormhole. Moreover, through pump-probe measurements, we experimentally measure the classical scalar field propagator at the outer edges of hyperbolic circuits to reproduce the two-point correlation function of a scalar operator on an equal-time slice of the dual 2D boundary CFT, and verify its exponential dependence on the boundary separation, with the scaling dimension associated with the scalar mass conforming to the Gubser-Polyakov-Klebanov-Witten (GPKW) formula in Lifshitz holography (*33-35*). Furthermore, using the two-point correlation function results, we reconstruct the entanglement entropy of a subsystem of the 2D boundary CFT and confirm its logarithmic scaling with the subsystem size, following the Ryu-Takayanagi (RT) formula (*36*). These observed behaviors of the holographic two-point correlation function and entanglement entropy exhibit excellent agreement with the theoretical predictions of the holographic duality, providing the first direct experimental evidence for a central tenet of holographic duality: the quantum properties of the boundary QFT can be quantitatively reproduced through the classical dynamics in its dual bulk gravity. Our work represents a crucial step toward establishing a new paradigm for investigating holographic models and the dynamics of negatively curved space through tabletop laboratory experiments.

**Results:**

We begin by considering the 3D Euclidean Lifshitz space in the large dynamical exponent limit, whose metric is expressed as: $\mathrm{d}s^2 = \mathrm{d}s_H^2 + \mathrm{d}\tau^2$ (see Materials and Methods S1), where $\mathrm{d}s_H^2$ denotes the metric of the 2D hyperbolic space and $\tau$ represents the Euclidean time. Particularly, since the $\tau$-coordinate is orthogonal to the hyperbolic space and its metric component is constant, the hyperbolic geometry remains invariant along the $\tau$ direction. Therefore, an effective way to model the 3D Euclidean Lifshitz space is to add the laboratory time as the $\tau$-coordinate onto the 2D hyperbolic space, thereby extending its dimensionality from two to three. In this work, we focus on two typical configurations of 3D Euclidean Lifshitz space with distinct topologies, i.e., pure Euclidean Lifshitz space (Fig. 1A) and Euclidean Lifshitz wormhole (Fig. 1B). To be more intuitive, their corresponding spatial geometries of a $\tau$-fixed slice can be represented as a two-sheet hyperboloid (Fig. 1C) and a one-sheet hyperboloid (Fig. 1D), respectively (*37*). These two geometries can be continuously projected as planar hyperbolic models, i.e., the "Poincaré disk" (Fig. 1E) and the "Poincaré ring" (Fig. 1F), respectively (see Supplementary Text S1.1 and S1.2), preserving the local constant negative curvature and global topology. More intriguingly, these two types of planar Poincaré models can be further discretized as type-I hyperbolic lattices (*14-32*) (Fig. 1G) and type-II hyperbolic lattices (Fig. 1H), respectively, via regular polygon tessellation. Furthermore, based on rigorous mapping between the tight-binding model and electric circuit network (*15-18*), these two types of hyperbolic lattices can be experimentally realized as type-I (Fig. 1I) and type-II hyperbolic lumped circuits (Fig. 1J), respectively. Consequently, by building on this top-down geometric consistency and introducing laboratory time as the third dimension, the physical properties of continuous classical scalar fields propagating in the 3D Euclidean Lifshitz space can be well-preserved through the evolution of discretized classical voltage fields in hyperbolic circuits along laboratory time, particularly in the long-wavelength regime (see Materials and Methods S3). In this setup, the dual boundary field is to a 2D quantum field, whose one dimension is defined along the boundary of the hyperbolic circuit and the other along the laboratory time.

In the experiment, we implement a typical finite type-I (type-II) hyperbolic circuit denoted by the Schläfli symbol $\{3, 7\}$ (extended Schläfli symbol $\{0.6533, 3, 7\}$), which represents a regular tessellation of the Poincaré disk (Poincaré ring with characteristic radius $R_H = 0.6533$) by 7 copies of 3-sided polygons meeting at each vertex (see Fig. S1). As illustrated by the hyperbolic circuits in Fig. 2, A and B, each node is grounded through an inductor $L$, and coupled with its seven nearest neighbors by seven capacitors $C$ (see zoomed-in view in Fig. 2A). Moreover, to implement the Dirichlet boundary conditions (*15*), all edge nodes are grounded by extra capacitors to ensure that each node is capacitively coupled to seven other nearest components. Note that although the physical distances between nodes vary, the on-node properties (e.g., resonant frequency) and inter-node couplings are identical across all nodes, ensuring a strict realization of hyperbolic lattices. This is justified by the fact that, in low-frequency lumped-element circuits, phase delays and parasitic effects originating from the wiring are negligible (*15-18*).

**Identification of spatiotemporal geodesics**

We begin our experiments by identifying spatiotemporal geodesics in the type-I and type-II hyperbolic circuits via time-resolved measurements (see Materials and Methods S4.2.1). As illustrated by the schematic experimental setups in Fig. 2, A and B, a temporal voltage pulse (green line in Fig. 2C) is fed into a type-I (type-II) hyperbolic circuit from edge node31 (red dot in Fig. 2A) (node300 (blue dot in Fig. 2B)). The corresponding frequency spectrum of the pulse (green line marked by the left-pointing arrow in Fig. 2D) is broad enough to cover the measured impedance spectra of type-I and type-II (red and blue lines marked by the right-pointing arrow in Fig. 2D) hyperbolic circuits (see Fig. S2 and Fig. S3), ensuring all intrinsic modes are excited to approximate the continuum response (*15*). The pulse propagating in hyperbolic circuits gives rise to variational instantaneous phases at nodes along the trajectory. We plot several instantaneous phase profiles, sampled at equal time intervals along the temporal evolution axis, for the type-I and type-II hyperbolic circuits, as shown in Fig. 2, E and F, respectively. At each time slice, the outermost wave front (black arcs in Fig. 2, E and F), delineated by the position of equal instantaneous phase, marks the leading edge of the pulse. Connecting these loci over time reveals the spatiotemporal geodesics (3D black arrows in Fig. 2, E and F) traced by the pulse in hyperbolic circuits. It can be seen that the spatiotemporal geodesics in the type-I hyperbolic circuit are 3D arc-shaped curves, whereas those in the type-II hyperbolic circuit form 3D bow-shaped curves that wrap around the "throat" of the wormhole (white dashed circles in Fig. 2F), owing to the non-trivial topology. Since the 3D bulk metric is time-independent, the projection of any bulk geodesic onto the time-fixed 2D hyperbolic subspace is itself a geodesic of that subspace (blue bottom planes in Fig. 2, E and F) (see Supplementary Text S1.3 and S2). These experimental results are precisely consistent with the theoretical results in the continuum (see Fig. S4).

**Reproducing quantum field behavior from classical dynamics in a type-I hyperbolic circuit**

Now we experimentally measure the scalar field propagators in hyperbolic circuits, using static pump-probe measurements (see Materials and Methods S4.2.4), to holographically extract the two-point correlation function and entanglement entropy of the dual QFT. We begin by considering the Poincaré disk setup with curvature radius-squared $\ell^2 = 1$, the spatial geometry of a $\tau$-fixed slice in the 3D pure Euclidean Lifshitz space that is dual to the 2D non-relativistic CFT. We first investigate the holographic two-point correlation function. The equal-time two-point correlation function $\langle \mathcal{O}(z_a, \tau_0)\mathcal{O}(z_b, \tau_0)\rangle$ of a scalar operator, inserted at 2D points $(z_a, \tau_0)$ and $(z_b, \tau_0)$ with 1D spatial coordinates $z_a = Re^{i\theta_a}$ and $z_b = Re^{i\theta_b}$($R = 1 - \epsilon$, $\epsilon \to 0$ is ultraviolet (UV) cutoff) in the boundary CFT, is conjectured by the holographic duality to follow (*33-35*) (see Supplementary Text S4.1):

$$\langle \mathcal{O}(z_a, \tau_0)\mathcal{O}(z_b, \tau_0)\rangle \propto e^{-\Delta \cdot L_{\gamma_{Ls}(z_a,\tau_0;z_b,\tau_0)}}, \quad (1)$$

where $L_{\gamma_{Ls}(z_a,\tau_0;z_b,\tau_0)}$ is the length of the geodesic connecting $(z_a, \tau_0)$ and $(z_b, \tau_0)$. Since the 2D Poincaré disk is a "totally geodesic submanifold" of the 3D pure Euclidean Lifshitz space (see Materials and Methods S1 and Supplementary Text S1.3), the geodesic $\gamma_{Ls}(z_a, \tau_0; z_b, \tau_0)$ coincides exactly with the geodesic $\gamma_{\mathcal{D}}(z_a, z_b)$ connecting $z_a$ and $z_b$ in the Poincaré disk, both having the length $L_{\gamma_{Ls}(z_a,\tau_0;z_b,\tau_0)} = L_{\gamma_{\mathcal{D}}(z_a,z_b)} = 2\log\left[\frac{2}{\epsilon}\sin\left(\frac{\theta_a-\theta_b}{2}\right)\right]$ (see

Supplementary Text S3.1). The conformal dimension $\Delta$ is related to the scalar mass $m_S$ via the GPKW formula in Lifshitz holography (*33-35*) (see Materials and Methods S1 and Supplementary Text S4.1):

$$\Delta(\Delta - d + 1) = m_S^2 \ell^2, \tag{2}$$

where the boundary dimension $d = 2$. Remarkably, upon discretizing the Poincaré disk into a type-I hyperbolic lattice, $\langle \mathcal{O}(z_a, \tau_0)\mathcal{O}(z_b, \tau_0)\rangle$ is found to be effectively reproduced by the lattice bulk-to-bulk propagator $G_l(z'_a, z'_b)$ between two near-boundary sites $z'_a = R' e^{i\theta_a}$ and $z'_b = R' e^{i\theta_b}$ (with $R' \to R$ and $L_{\gamma_{\mathcal{D}}(z_a, z_b)} \approx L_{\gamma_{\mathcal{D}}(z'_a, z'_b)}$) of the discretized weakly coupled scalar field (see Supplementary Text S4.3), up to an undetermined overall factor $\alpha$, such that $\langle \mathcal{O}(z_a, \tau_0)\mathcal{O}(z_b, \tau_0)\rangle = \alpha G_l(z'_a, z'_b)$. Since this factor does not alter the scaling behavior described by Eq. 2 and the retrieval of the scaling exponent, here we set $\alpha = 1$. This key point allows for testing the holographic conclusions of Eq. 1 and Eq. 2 by reproducing $\langle \mathcal{O}(z_a, \tau_0)\mathcal{O}(z_b, \tau_0)\rangle$ through experimentally measuring $G_l(z'_a, z'_b)$ in a type-I hyperbolic circuit. In the experiment, we extract the mass-dependent $G_l(z'_a, z'_b; m_S)$ by measuring its circuit equivalent, i.e., the frequency-dependent reduced circuit Green's function $G_{c_R}(z'_a, z'_b; \omega)$ of the voltage field in the type-I hyperbolic circuit, where $m_S$ is determined by the circuit frequency $\omega$ via the relation $m_S^2 = -4/(7\omega^2 LC)$ (see Materials and Methods S3). To best approximate the near-boundary limit ($R' \to R$), we extract $G_l(z'_a, z'_b)$ from the outmost edge nodes. As shown by the schematic experimental setup in Fig. 3A, we fix the first test point (labelled by $z'_a$) and vary the second test point (labelled by $z'_b$) along the edge to change the node-node separation. In stable field consideration, we set input frequency $f = \omega/2\pi$ above the resonance frequency of the fundamental mode (see Fig. S2), ensuring $m_S^2$ is larger than the Breitenlohner-Freedman bound (*25*). As shown in Fig. 3B, for three different $m_S^2$, the experimentally reproduced $\langle \mathcal{O}(z_a, \tau_0)\mathcal{O}(z_b, \tau_0)\rangle$ (color dots) exhibits a clear exponential decay with respect to $L_{\gamma_{LS}(z_a, \tau_0; z_b, \tau_0)}$, as evidenced by the well-fitted solid lines conforming to Eq. 1. This behavior is further confirmed by the linear relation of the logarithmic plot shown in the inset. Moreover, we also plot the retrieved scaling dimension $\Delta$ (red dots) as a function of $m_S^2$, as shown in Fig. 3C. The red line indicates the best fit to formula $\Delta(\Delta - d_l + 1) = m_S^2 {\ell^2}_l$ with a lattice curvature radius-squared ${\ell^2}_l = 0.6257$ and a lattice boundary dimension $d_l = 2.1849$, verifying the Lifshitz GPKW formula. The inset displays a linear relation between $\Delta(\Delta - d_l + 1)$ and $m_S^2$ with a slope of ${\ell^2}_l$. Note that the fitted ${\ell^2}_l$ and $d_l$ slightly deviate from the expected values of 0.84 (*22, 23*) and 2 in lattice model, primarily due to the insufficient boundary discretization in the fabricated type-I hyperbolic circuit. These deviations can be effectively eliminated by increasing the lattice radius (see Fig. S5).

We then proceed to investigate holographic entanglement entropy. Holographic duality conjectures that the average entanglement entropy $\overline{S(A)}$ over central charge $c$ for an interval $A$ as a subsystem of the boundary CFT follows the RT formula (*36*):

$$\overline{S(A)} = \frac{S(A)}{c} = \frac{1}{6} L_{\gamma_{LS}(A)} = \frac{1}{3}\log\left[\frac{2}{\epsilon}\sin\left(\frac{\theta_A}{2}\right)\right] = \frac{1}{3}\log\left[2\sin\left(\frac{\theta_A}{2}\right)\right] + \sigma, \tag{3}$$

where $c = 3/2G_N^{(3)}$, $G_N^{(3)}$ is the 3D Newton constant, $L_{\gamma_{LS}(A)}$ is the area (i.e., length) of a static minimal surface (i.e., geodesic) $\gamma_{LS}(A)$ anchored at the ends of $A$, $\sigma = -\frac{1}{3}\log\epsilon$ is the UV intercept. Note that this formula captures the leading classical contribution in bulk gravity corresponding to the limit $G_N^{(3)} \to 0$, which in turn implies that the dual boundary CFT has $c \to \infty$. Remarkably, Eq.

3 is applicable to a broad class of CFTs in the large $c$ limit. Here, we focus on a scalar Gaussian conformal field with $\mathrm{O}(c)$ symmetry, a well-established holographic model known as “generalized free field” (*10, 11, 38-40*), whose dual bulk gravity contains $c$-copies of a free Klein-Gordon field. In this setup, the generating functional of the boundary CFT factorizes into a direct product of $c$ copies of the generating functional of a single boundary scalar field (see Supplementary Text S5.4). On the other hand, the ground state wave functional of a boundary scalar field can be derived from its generating functional, which is dual to the counterpart of a bulk free scalar field that approximates a Gaussian form (see Supplementary Text S5.3.1). These key points enable us to reconstruct the holographic entanglement entropy as $S(A) = c\overline{S(A)}$, where $\overline{S(A)}$ is the entanglement entropy contributed by a single scalar field and can be fully determined by correlation function $\langle \mathcal{O}(z_a, \tau_0)\mathcal{O}(z_b, \tau_0)\rangle$. Upon discretizing the Poincaré disk into the type-I hyperbolic lattice, $\overline{S(A)}$ of an $n$-site subsystem $A$ of the $N$-site boundary system can be computed via equation (see Supplementary Text S5.3.1):

$$\overline{S(A)} = \mathrm{tr}\left[\left(C_A + \frac{1}{2}\right)\ln\left(C_A + \frac{1}{2}\right) - \left(C_A - \frac{1}{2}\right)\ln\left(C_A - \frac{1}{2}\right)\right], \tag{4}$$

Where $C_A = \sqrt{X_{A,\varepsilon}P_{A,\varepsilon}}$, $X_{A,\varepsilon}$ ($P_{A,\varepsilon}$) is the $n \times n$ block of the $N \times N$ matrix $X_\varepsilon = \varepsilon^{-1}\mathbb{J} + X$ ($P_\varepsilon = X_\varepsilon{}^{-1}/4$) with elements restricted to $A$, $\mathbb{J}$ is the all-one matrix, $\varepsilon$ is a small regulator, and the elements of $X$ consist of $\langle \mathcal{O}(z_a, \tau_0)\mathcal{O}(z_b, \tau_0)\rangle$ between arbitrary boundary sites. Note that without $\varepsilon^{-1}J$ term, Eq. 4 cannot be directly used to calculate $\overline{S(A)}$ since the boundary CFT has scaling symmetry (see Supplementary Text S5.2 and S5.3). We henceforth set $\varepsilon = 1/64$, which is sufficiently small to ensure that the computed $\overline{S(A)}$ converges up to a constant offset (see Fig. S6). This offset can be absorbed into the UV cutoff $\epsilon$ and therefore does not affect the scaling behavior of $\overline{S(A)}$. Additionally, we have experimentally demonstrated that $\langle \mathcal{O}(z_a, \tau_0)\mathcal{O}(z_b, \tau_0)\rangle$ can be effectively reproduced by $G_l(z'_a, z'_b)$. The overall factor $\alpha$ between them does not affect $\overline{S(A)}$, because $C_A$ is independent of $\alpha$ as $X_{A,\varepsilon}P_{A,\varepsilon}$ is scale-invariant. Therefore, we also set $\alpha = 1$. Consequently, we can test the holographic conclusion of Eq. 3 by reconstructing $\overline{S(A)}$ through measuring the correlation matrix $X_\varepsilon$. In the experiment, we measure $G_l(z'_a, z'_b)$ with $m_S^2 = -0.2413$ across all edge nodes, as illustrated in Fig. 3D, to construct a full $X$ shown in Fig. 3E. Here, site/node indexes are assigned based on their polar angles, as labeled in the right panel of Fig. 3D. Note that the long-range two-point correlation function beyond the fourth-nearest-neighbor nodes are omitted in the measurements (see Materials and Methods S4.2.4). We plot the reconstructed $\overline{S(A)}$, averaged over 98 different $X_\varepsilon$ constructed by successively shifting the starting edge site backward, as a function of $\theta_A$ that quantifies the size of $A$, as shown in Fig. 3F. As $\theta_A$ increases, $\overline{S(A)}$ initially grows logarithmically, then asymptotically tends to a constant as $\theta_A$ approaches $\pi$. Beyond this point, $\overline{S(A)}$ decreases symmetrically, eventually vanishes as $\theta_A$ approaches $2\pi$. Obviously, the relation $\overline{S(A)} = \overline{S(\bar{A})}$ ($\bar{A}$ is the complement of $A$) holds universally for all $\theta_A$, as the total system $A \cup \bar{A}$ is in a pure state. This behavior is precisely consistent with the theoretical law of Eq. 3, as evidenced by the fitted blue line with an effective UV intercept $\sigma = 4.169$ in Fig. 3F, confirming the RT formula. The inset further affirms the linear relation between $\overline{S(A)}$ and $L_{\gamma_{Ls}(A)}$ with a slope of $1/6$. Note that for different $m_S^2$, $\overline{S(A)}$ remains unchanged (see Fig. S7). Moreover, a finer boundary discretization in the type-I hyperbolic circuit would refine the result to precisely match the theoretical predictions (see Fig. S5).

**Reproducing quantum field behavior from the classical field in a type-II hyperbolic circuit**

Next, we consider the Poincaré ring setup with curvature radius-squared $\ell^2 = 1$, the spatial geometry of a $\tau$-fixed slice in the Euclidean Lifshitz wormhole that is dual to the 2D non-relativistic CFT. Although the global geometric topology of this space is different from that of the Poincaré disk, the holographic duality predicts that the equal-time two-point correlation function $\langle \mathcal{O}(z_a, \tau_0)\mathcal{O}(z_b, \tau_0)\rangle$ of the outer boundary CFT follows a similar rule as Eq. 1 (*33-35*) (see Supplementary Text S4.1):

$$\langle \mathcal{O}(z_a, \tau_0)\mathcal{O}(z_b, \tau_0)\rangle \propto e^{-\Delta \cdot L_{\gamma_{Lw}(z_a,\tau_0;z_b,\tau_0)}}, \tag{5}$$

where $L_{\gamma_{Lw}(z_a,\tau_0;z_b,\tau_0)}$ is the length of the geodesic connecting points $(z_a, \tau_0)$ and $(z_b, \tau_0)$. Since the 2D Poincaré ring is a "totally geodesic submanifold" of the 3D Euclidean Lifshitz wormhole (see Materials and Methods S1 and Supplementary Text S1.3), the geodesic $L_{\gamma_{Lw}(z_a,\tau_0;z_b,\tau_0)}$ coincides exactly with the geodesic $\gamma_{\mathcal{R}}(z_a, z_b)$ connecting points $z_a$ and $z_b$ in the Poincaré ring, both having the length $L_{\gamma_{Lw}(z_a,\tau_0;z_b,\tau_0)} = L_{\gamma_{\mathcal{R}}(z_a,z_b)} = 2\log\left[\frac{\beta}{\pi\epsilon}\sinh\left(\frac{\pi(\theta_a-\theta_b)}{\beta}\right)\right]$ (see Supplementary Text S3.2). Here $\beta = 2\pi/r_h$ and $r_h = -\pi/(2\ln R_H)$ is the "throat" radius of the wormhole. The conformal dimension $\Delta$ continues to associate with the scalar mass $m_S$ via the Lifshitz GPKW formula in Eq. 2. To test these holographic conclusions, we follow the aforementioned strategy by experimentally measuring the scalar field propagator $G_l(z'_a, z'_b)$ as a function of $m_S^2$ between two nodes with varying separations along the outer edge of a type-II hyperbolic circuit, thereby reproducing $\langle \mathcal{O}(z_a, \tau_0)\mathcal{O}(z_b, \tau_0)\rangle$. As shown in Fig. 4, B and C, the reproduced $\langle \mathcal{O}(z_a, \tau_0)\mathcal{O}(z_b, \tau_0)\rangle$ showcases an exponential decay with $L_{\gamma_{Lw}(z_a,\tau_0;z_b,\tau_0)}$, conforming to Eq. 5. This result indicates that the dual boundary CFT is in a thermal state of temperature $T = 1/\beta$. Besides, the retrieved scaling dimension $\Delta$ is found to relate to $m_S$ via $\Delta(\Delta - d_l + 1) = m_S^2 {\ell^2}_l$ with a lattice curvature radius-squared ${\ell^2}_l = 0.603$ and a lattice boundary dimension $d_l = 2.1747$, verifying the Lifshitz GPKW formula. Note that these values also can be optimized to the expected values of 0.84 (*22, 23*) and 2 in lattice model by increasing the outer lattice radius (see Fig. S8).

We then proceed to investigate the holographic entanglement entropy. In the high temperature limit $\beta/L_b = r_h^{-1} \ll 1 (L_b = 2\pi\ell = 2\pi)$, holographic duality predicts the average entanglement entropy over different degrees of freedom $\overline{S(A)}$ for an interval $A$ as a subsystem of the outer boundary CFT follows the RT formula (*36*):

$$\overline{S(A)} = \frac{S(A)}{c} = \frac{1}{6}L_{\gamma_{Lw}(A)} = \frac{1}{3}\log\left[\frac{\beta}{\pi}\sinh\left(\frac{\pi\theta_A}{\beta}\right)\right] + \sigma, \tag{6}$$

where $L_{\gamma_{Lw}(A)}$ is the area (i.e., length) of a static minimal surface (i.e., geodesic) $\gamma_{Lw}(A)$ anchored at the ends of $A$. Since the outer boundary CFT is in a thermal state of temperature $T$, its dual bulk gravity possesses a Gaussian thermal state, enabling $\overline{S(A)}$ to be fully reconstructed by one scalar field correlation function $\langle \mathcal{O}(z_a, \tau_0)\mathcal{O}(z_b, \tau_0)\rangle$. Consequently, after discretizing the Poincaré ring into type-II hyperbolic lattice, $\overline{S(A)}$ of an $n$-site outer boundary subsystem $A$ can also be computed via Eq. 4, but with the revised matrixes $X_\varepsilon = V^{-1}\coth\beta V$ and $P_\varepsilon = V\coth\beta V$, where $V$ is a symmetric matrix (see Supplementary Text S5.3.3). To reconstruct $\overline{S(A)}$ to test the holographic conclusion of Eq. 6, following the aforementioned strategy, we need to experimentally reproduce$\langle \mathcal{O}(z_a, \tau_0)\mathcal{O}(z_b, \tau_0)\rangle$ to construct a correlation matrix $X_\varepsilon$. Note that in this case, the overall factor $\alpha$ between $\langle \mathcal{O}(z_a, \tau_0)\mathcal{O}(z_b, \tau_0)\rangle$ and $G_l(z'_a, z'_b)$ markedly affects $\overline{S(A)}$ since $C_A$ is dependent of $\alpha$ as $X_{A,\varepsilon}P_{A,\varepsilon}$ is not scale-invariant. Therefore, we treat $\alpha$ as a fit parameter. In the experiment, we measure $G_l(z'_a, z'_b)$ with $m_S^2 = -0.2413$ across all outer edge nodes in a type-II hyperbolic lattice, as illustrated in Fig. 4D, to construct a full correlation matrix $X$ shown in Fig. 4E. The long-range correlations beyond the fourth-nearest-neighbor nodes are also omitted in the measurements (see Materials and Methods S4.2.4). We plot the reconstructed $\overline{S(A)}$ with best fit to $\alpha = 120$, averaged over 144 different $X_\varepsilon$ constructed by successively shifting the starting edge site backward, as a function of $\theta_A$, as shown in Fig. 4F. As $\theta_A$ increases, $\overline{S(A)}$ initially grows logarithmically and then continuous to increase asymptotically as $\theta_A$ approaches $2\pi$. Obviously, the relation $\overline{S(A)} = \overline{S(\bar{A})}$ no longer holds, indicating that the total system $A \cup \bar{A}$ resides in a thermal state. This behavior is basically consistent with the theoretical law of Eq. 6, as indicated by the fitted blue line with an effective UV intercept $\sigma = 5.034$ in Fig. 4F, confirming the RT formula. The inset further affirms the linear relation between $\overline{S(A)}$ and $L_{\gamma_{Lw}(A)}$ with a slope of $1/6$. Note that for different $m_S^2$, $\overline{S(A)}$ nearly remain unchanged (see Fig. S7), Moreover, a finer outer boundary discretization in type-II hyperbolic circuit would refine the result to more precisely match the theoretical predictions (see Fig. S8).

We note that a deviation from the theoretical prediction emerges in the large $\theta_A$ regime, primarily due to the insufficiently large $R_H = 0.6533$ (i.e., $r_h^{-1} = 0.271$) of our fabricated type-II hyperbolic circuit, which does not fully satisfy the high temperature limit $r_h^{-1} \ll 1$. Such deviations can be effectively mitigated by increasing $R_H$. As illustrated in Fig. 5, A and C, we experimentally implement another type-II hyperbolic circuit with a larger $R_H = 0.8677$ (i.e., $r_h^{-1} = 0.0903$), the deviation of reconstructed $\overline{S(A)}$ in the large $\theta_A$ regime is clearly diminished, as shown in Fig. 5D. Moreover, further increasing $R_H$ eliminates the deviation entirely, as evidenced by the combined experimental-numerical result in Fig. 5E for a type-II hyperbolic circuit with an extremely large $R_H = 0.9652$ (i.e., $r_h^{-1} = 0.0225$) (see Materials and Methods S4.2.4). Note that the factor $\alpha$ is a type-II hyperbolic lattice-dependent parameter. Specifically, for a fixed $\{p,\ q\}$ and number of layers, it can be proved that $\alpha$ depends on the temperature $T$ according to $\alpha = \alpha_0\, (4\pi T)^{2\Delta-1}$ with a temperature-independent factor $\alpha_0$ (see Supplementary Text S5.3.4 and Fig. S9).

**Discussions:**

In this work, we present the first experimental test of the celebrated holographic duality between a 3D bulk gravity and a 2D QFT using type-I and type-II hyperbolic circuits, which emulate the pure 3D Euclidean Lifshitz space and the 3D Euclidean Lifshitz wormhole, respectively, in the large dynamical exponent limit, with laboratory time acting as the third dimension. We experimentally identify distinct geodesic behaviors of 3D Euclidean Lifshitz spaces in the presence or absence of an analog wormhole. Furthermore, by experimentally measuring the classical scalar field propagator at the outer edges of hyperbolic circuits, we extract the equal-time two-point correlation function and entanglement entropy of the dual 2D boundary CFT, corroborating that these two pivotal holographic quantities indeed reveal behaviors congruent with the predictions of the holographic duality. These results thereby provide the first direct experimental evidence for a central tenet of holographic duality: the quantum properties of the boundary QFT can be quantitatively reproduced through the classical dynamics in its dual bulk gravity. This work not only expands the fundamental research scope of hyperbolic lattice physics but also provides a new strategy for mimicking novel quantum-gravity-inspired phenomena in classical systems with negatively curved spaces. Moreover, this versatile platform possesses long-term viability owing to its flexible hyperbolic lattice layout, which avoids the limitation of boundary site congestion (see Fig. S10). This advantage also offers promising opportunities to explore unorthodox and intricate spacetime geometry, such as the replica wormhole (*41, 42*) and time-evolving curved space (*43*). Straightforward upgrades will enable us to generalize hyperbolic lattices to quantum models by adopting superconducting circuits and quantum qubits to access the real quantum physical effects, such as the "ER=EPR" conjecture (*44*) and quantum information scrambling (*45*) in negatively curved spaces.

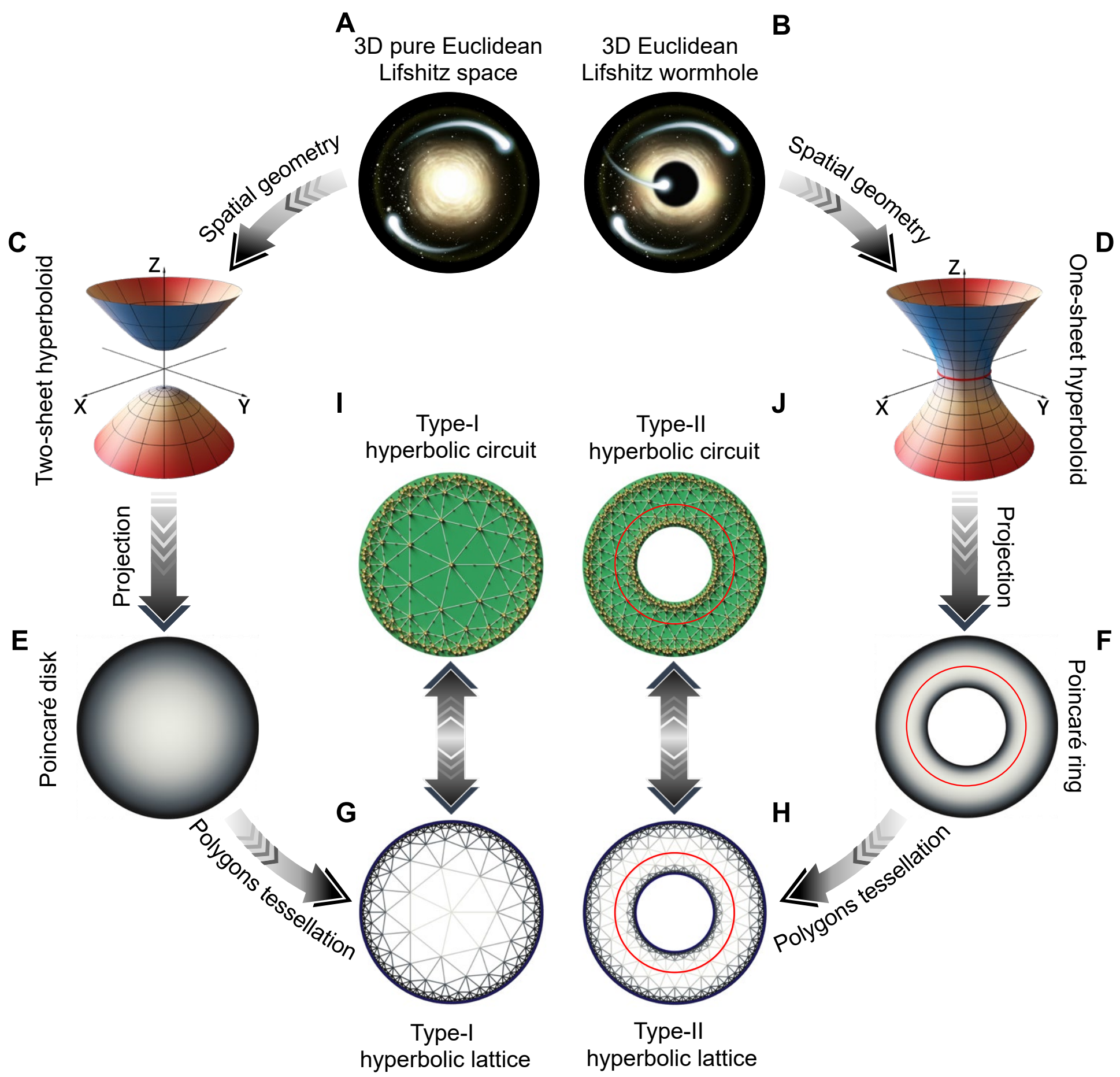

**Fig. 1. Correspondence between 3D Euclidean Lifshitz spaces and hyperbolic circuits.** (**A** and **B**) Schematic illustrations of the 3D pure Euclidean Lifshitz space (A) and the 3D Euclidean Lifshitz wormhole (B). (**C** and **D**) Two-sheet (C) and one-sheet (D) hyperboloids, which are spatial geometries of (A) and (B), respectively. The red circle in (D) indicates the location with the minimum radius on the surface, i.e., the "throat" of the wormhole. (**E** and **F**) Poincaré disk (E) and ring (F) models, which are projective planar counterparts of (C) and (D), respectively. (**G** and **H**) Type-I (G) and type-II (H) hyperbolic lattices, which are lattice regularizations of (E) and (F), respectively. (**I** and **J**) Type-I (I) and type-II (J) hyperbolic circuits, which are circuit equivalents of (G) and (H), respectively. The red circle in (F), (H), and (J) is the exact counterpart of the "throat" of the wormhole in (D).

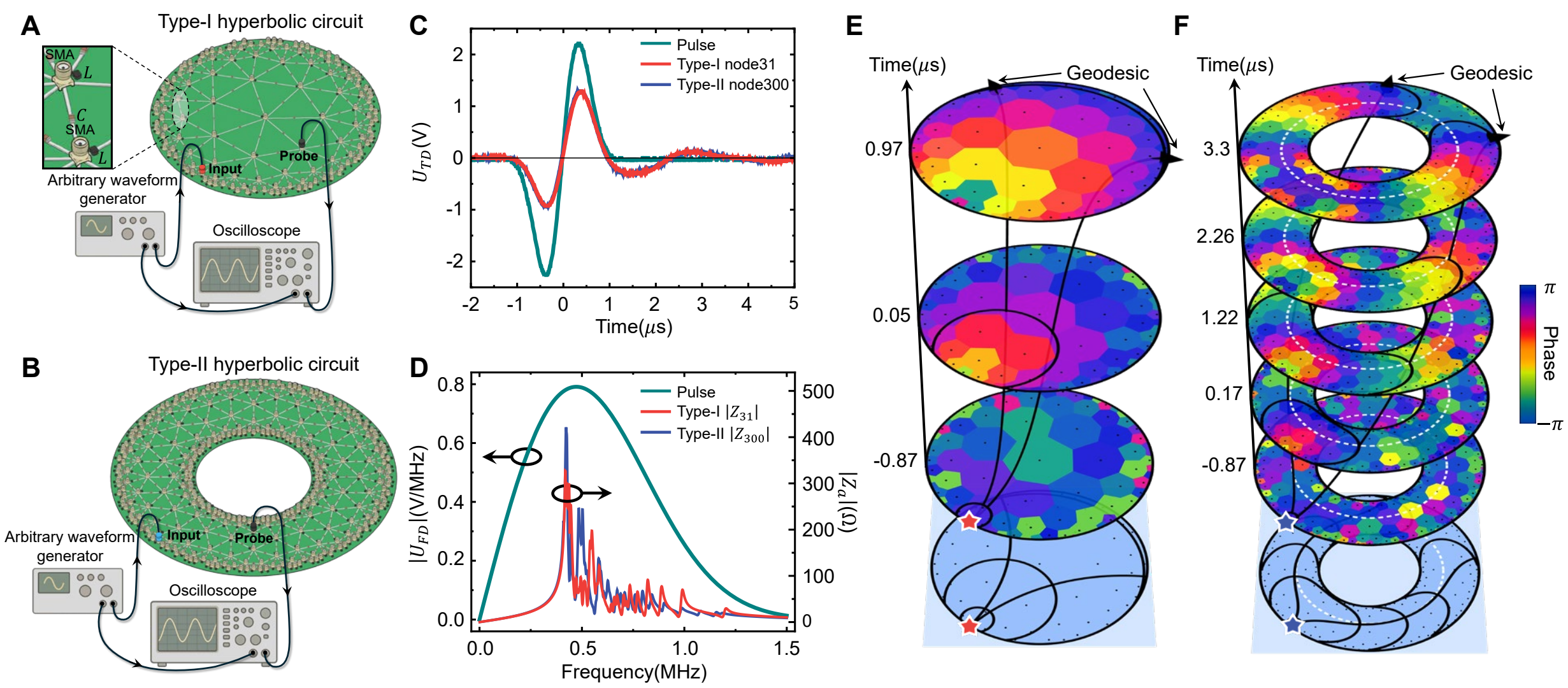

**Fig. 2. Identification of spatiotemporal geodesics in type-I and type-II hyperbolic circuits.** (**A** and **B**) Schematic illustrations of the experimental setups for time-resolved measurement in a {3, 7} type-I hyperbolic circuit (A) and a {0.6533, 3, 7} type-II hyperbolic circuit (B). The inset in (A) shows the zoomed-in view of the circuit element configurations of two connected nodes. (**C**) Measured voltage excitation pulse (green line) and voltage response pulses (red and blue lines) at the input nodes. (**D**) Measured frequency spectrum of the excitation pulse (green line plotted with respect to the left axis, marked by the left-pointing arrow) and the impedances to ground at the input nodes (red and blue lines plotted with respect to the right axis, marked by the right-pointing arrow). (**E** and **F**) Measured instantaneous phase profiles (color map) of the pulse propagating in type-I (E) and type-II (F) hyperbolic circuits, sampled at equal time intervals. In each time slice, small black dots represent the nodes, black arc represents the outermost wave front, and white dashed circle represents the "throat" of the wormhole. The 3D black arrows indicate representative geodesics. The bottom blue planes depict the projections of wave fronts and geodesics.

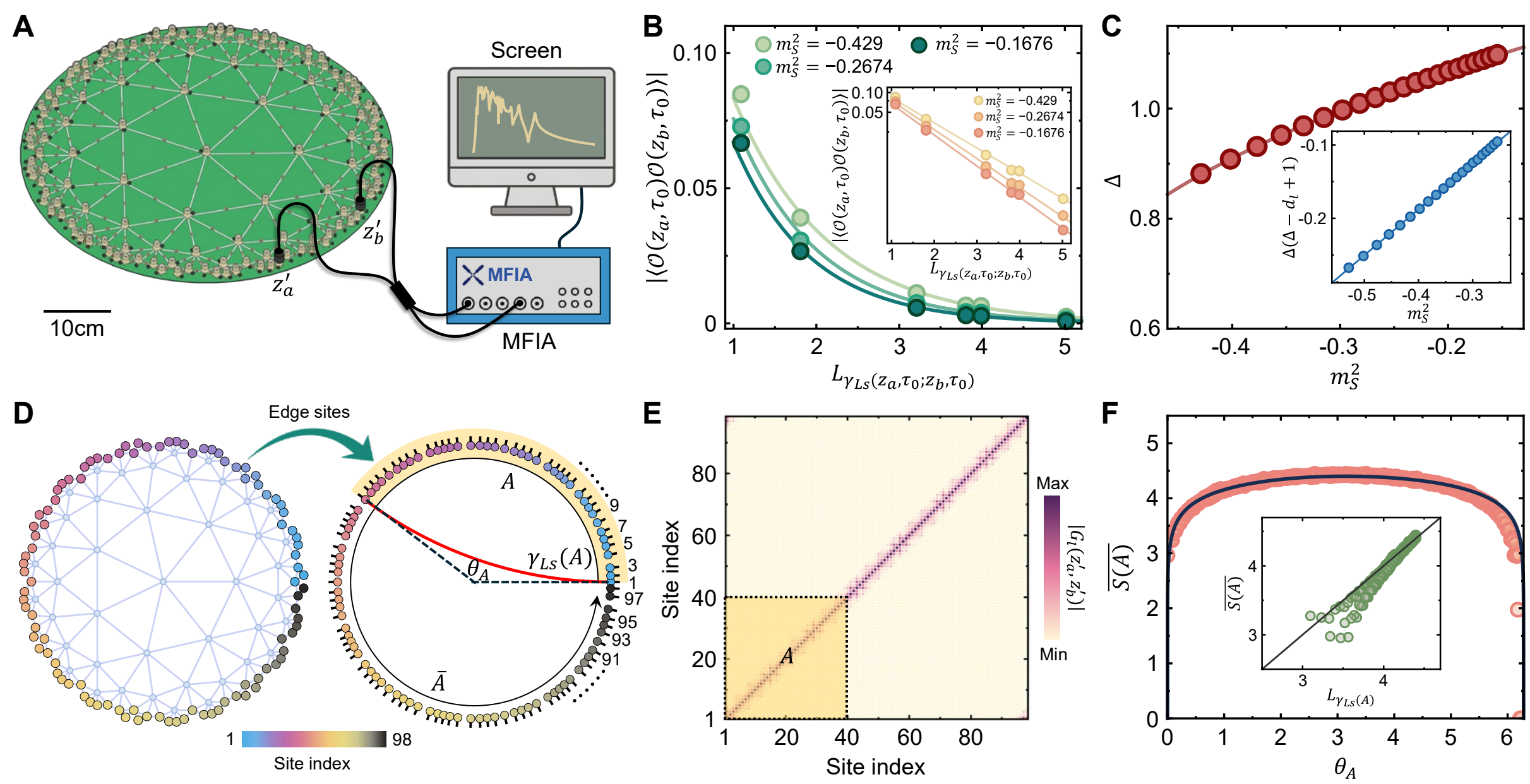

**Fig. 3. Extraction of holographic two-point correlation function and entanglement entropy in a type-I hyperbolic circuit.** (**A**) Schematic illustration of the experimental setup for measuring the scalar field propagator $G_l(z_a', z_b')$ in a {3, 7} type-I hyperbolic circuit. (**B**) Reproduced two-point correlation function $\langle \mathcal{O}(z_a, \tau_0)\mathcal{O}(z_b, \tau_0)\rangle$ (color dots) versus geodesic length $L_{\gamma_{Ls}(z_a,\tau_0;z_b,\tau_0)}$ for three different mass-squareds $m_S^2$. The color solid line is a theoretical fit to Eq.1. The inset depicts $\langle \mathcal{O}(z_a, \tau_0)\mathcal{O}(z_b, \tau_0)\rangle$ as a linear function of $L_{\gamma_{Ls}(z_a,\tau_0;z_b,\tau_0)}$ on a logarithmic scale. (**C**) Retrieved conformal dimension $\Delta$ (red dots) versus $m_S^2$. The red solid line shows the theoretical fit to the formula $\Delta(\Delta - d_l + 1) = m_S^2 \ell_l^2$. The inset displays $\Delta(\Delta - d_l + 1)$ as a linear function of $m_S^2$. (**D**) Schematic illustration of building up the edge sites system from a type-I hyperbolic lattice. The orange region represents a subsystem $A$ whose size is quantified by the angle $\theta_A$, the red arc represents the anchored geodesic $\gamma_{Ls}(A)$. All edge sites are sequentially labeled in a counterclockwise direction. (**E**) Reconstructed correlation matrix $X$ with $m_S^2 = -0.2413$ for the entire edge system. The orange region marks the submatrix corresponding to $A$ in (D). (**F**) Reconstructed entanglement entropy $\overline{S(A)}$ (pink dots) versus $\theta_A$. The dark blue line shows the theoretical fit to Eq. 3. The inset depicts $\overline{S(A)}$ as a linear function of $L_{\gamma_{Ls}(A)}$. The $\overline{S(A)}$ data is averaged over 98 different constructions of $X$, binned in intervals of 0.02 rad.

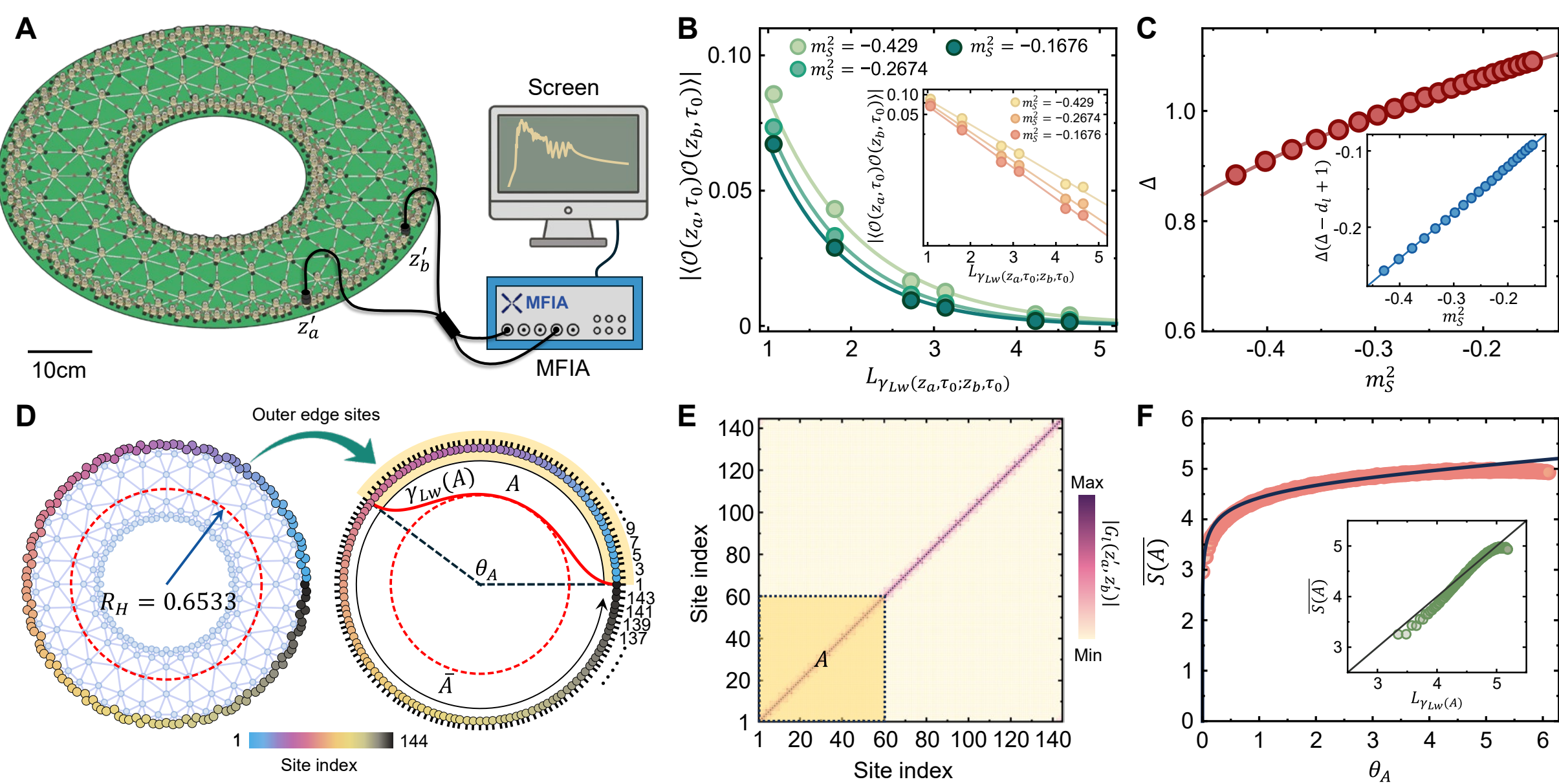

**Fig. 4. Extraction of holographic two-point correlation function and entanglement entropy in a type-II hyperbolic circuit.** (**A**) Schematic illustration of the experimental setup for measuring the scalar field propagator $G_l(z'_a, z'_b)$ in a $\{0.6533, 3, 7\}$ type-II hyperbolic circuit. (**B**) Reproduced two-point correlation function $\langle \mathcal{O}(z_a, \tau_0)\mathcal{O}(z_b, \tau_0)\rangle$ (color dots) versus geodesic length $L_{\gamma_{Lw}(z_a,\tau_0;z_b,\tau_0)}$ for three different mass-squareds $m_S^2$. The color solid line is a theoretical fit to Eq. 4. The inset depicts $\langle \mathcal{O}(z_a, \tau_0)\mathcal{O}(z_b, \tau_0)\rangle$ as a linear function of $L_{\gamma_{Lw}(z_a,\tau_0;z_b,\tau_0)}$ on a logarithmic scale. (**C**) Retrieved conformal dimension $\Delta$ (red dots) versus $m_S^2$. The red solid line shows the theoretical fit to the formula $\Delta(\Delta - d_l + 1) = m_S^2 \ell_l^2$. The inset displays $\Delta(\Delta - d_l + 1)$ as a linear function of $m_S^2$. (**D**) Schematic illustration of building up the outer edge sites system from a type-II hyperbolic lattice. The orange region represents a subsystem $A$ whose size is quantified by the angle $\theta_A$, the red arc indicates the anchored geodesic $\gamma_{Lw}(A)$. All edge sites are sequentially labeled in a counterclockwise direction. (**E**) Reconstructed correlation matrix $X$ with $m_S^2 = -0.2413$ for the entire outer edge system. The orange region marks the submatrix corresponding to $A$ in (D). (**F**) Reconstructed entanglement entropy $\overline{S(A)}$ (pink dots) versus $\theta_A$. The dark blue line shows the theoretical fit to Eq. 6. The inset depicts $\overline{S(A)}$ as a linear function of $L_{\gamma_{Lw}(A)}$. The $\overline{S(A)}$ data is averaged over 144 different constructions of $X$, binned in intervals of 0.02 rad.

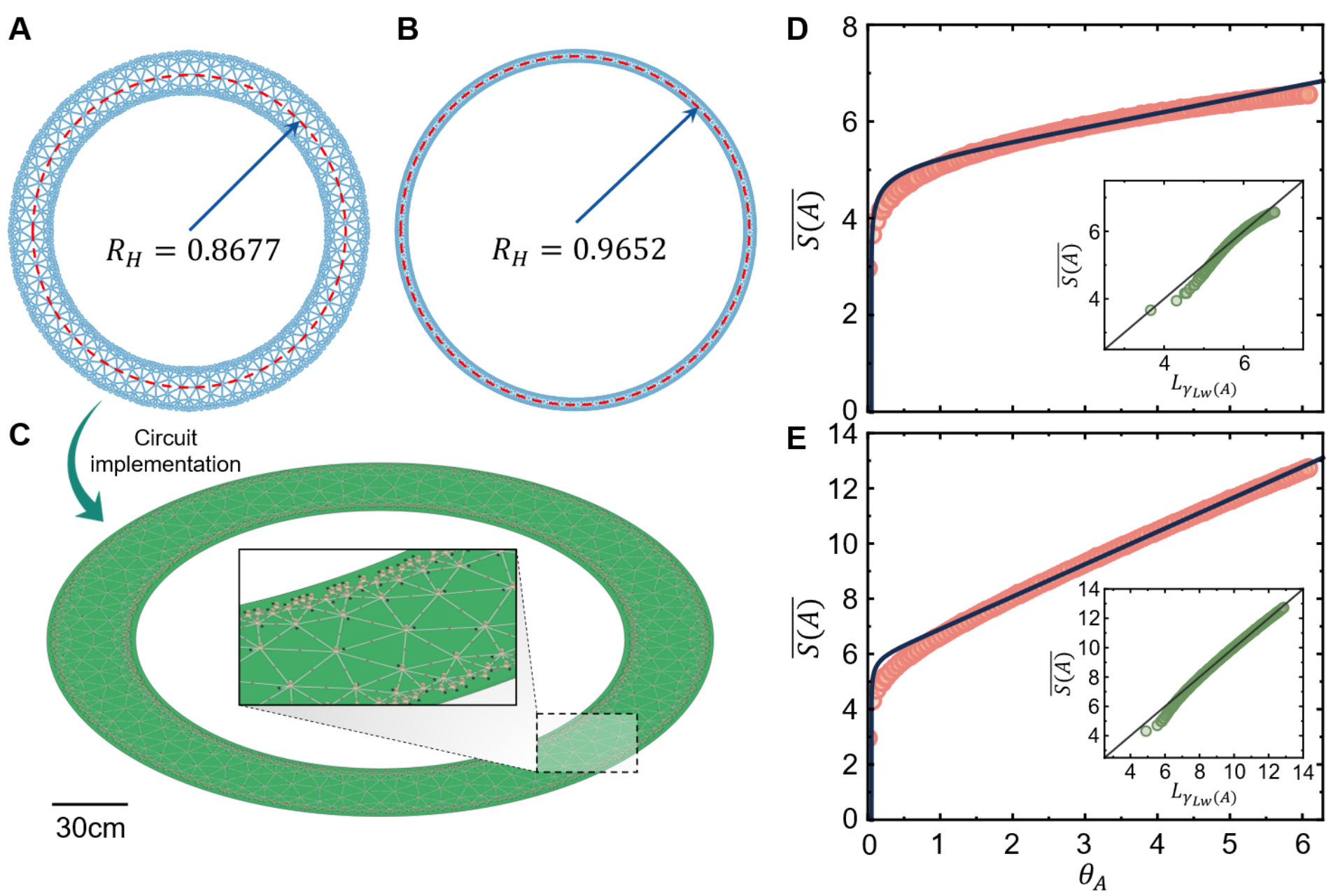

**Fig. 5. Extraction of holographic entanglement entropy in type-II hyperbolic circuits with larger radii.** (**A** and **B**) {0.8677, 3, 7} (A) and {0.9652, 3, 7} (B) type-II hyperbolic lattices. (**C**) Circuit implementation of (A). The inset shows the zoomed-in view of the circuit. (**D** and **E**) Experimentally (D) and numerically (E) reconstructed entanglement entropies $\overline{S(A)}$ (pink dots) versus angle $\theta_A$ with $m_S^2 = -0.2413$ for the type-II hyperbolic circuits corresponding to the lattices in (A) and (B), respectively. The dark blue line is a theoretical fit to Eq. 6. The inset depicts $\overline{S(A)}$ as a linear function of $L_{\gamma_{Lw}(A)}$. The $\overline{S(A)}$ data are averaged over 432 and 1728 different constructions of correlation matrix $X$, respectively, binned in intervals of 0.02 rad. The fitted factor $\alpha$ and UV intercept $\sigma$ are $\alpha = 340$, $\sigma = 5.82$, and $\alpha = 1290$, $\sigma = 7.05$, respectively.

**Acknowledgments:**

**Funding:** Z.G. acknowledges funding from the National Key R&D Program of China (2025YFA1412300), the National Natural Science Foundation of China (62361166627 and 62375118), Guangdong Basic and Applied Basic Research Foundation (2024A1515012770), Shenzhen Science and Technology Innovation Commission (202308073000209), and High-level Special Funds (G03034K004). Y.H.Y acknowledges funding from the Key Research and Development Program of the Ministry of Science and Technology (grant no. 2022YFA1404902 and 2022YFA1405201), and the National Natural Science Foundation of China (grant no. 62175215). R.-Q.Y. acknowledges funding from the National Natural Science Foundation of China (grant no. 12005155). Y.T.Y acknowledges funding from the Natural Science Foundation of Jiangsu Province (grant no. BK20200630) and the National Natural Science Foundation of China (grant no. 12004425). R.-G.C. acknowledges funding from the National Natural Science Foundation of China (grant no. 11821505).

**Author contributions:**

Z.G. and Y.H.Y. initiated the project. J.C., F.C., and R.-Q.Y. performed the theoretical calculation. Z.G., J.C., and Y.T.Y. designed the experiments. J.C. and Z.G. fabricated the samples. J.C. and Z. C. carried out the measurements. J.C., L.Y., Y.M., B.Y., X.X., X.Z., G.-G.L., P.P.S., H.C., R.-G.C., R.-Q.Y., Y.H.Y., and Z.G. analyzed the results. J.C. and Z.G. wrote the manuscript with input from all authors. Z.G., J.C., Y.H.Y., R.-Q.Y., and R.-G.C. revised the manuscript. Z.G., Y.H.Y., and R.-Q.Y. supervised the project.

**Competing interests:** The authors declare that they have no competing interests.

**Data and materials availability:** All data that support the plots within this paper and other findings of this study are available from the corresponding authors upon reasonable request.